\documentclass{PoS}
\usepackage{amsmath,amssymb}
\def\beq{\begin{equation}}
\def\eeq{\end{equation}}
\def\bea{\begin{eqnarray}}
\def\eea{\end{eqnarray}}
\def\BAN{\begin{eqnarray*}}
\def\EAN{\end{eqnarray*}}

\def\tr{\mathrm{tr}}

\def\g5{\gamma_5}

\newcommand{\Dodwf}{\mathcal{D}}

\title{Lattice QCD with Optimal Domain-Wall Fermion: Light Meson Spectroscopy}

\ShortTitle{LQCD with ODWF: Light Meson Spectroscopy}

\author{TWQCD Collaboration:
    Yu-Chih~Chen$^{1}$,
    \speaker{Ting-Wai~Chiu}$^{,1,2}$\thanks{Email: twchiu@phys.ntu.edu.tw}, 
    Tian-Shin~Guu$^{3}$, 
    Tung-Han~Hsieh$^{4}$,
    Chao-Hsi~Huang$^3$,
    Yao-Yuan~Mao$^1$ 
    \\
    $^1$ Department of Physics, and Center for Theoretical Sciences, National Taiwan University, Taipei 10617, Taiwan \\
    $^2$ Center for Quantum Science and Engineering, National Taiwan University, Taipei 10617, Taiwan \\
    $^3$ Center for General Education, and Institute of CSIE, National ILan University, I-Lan 260, Taiwan \\ 
    $^4$ Research Center for Applied Sciences, Academia Sinica, Taipei 115, Taiwan}

\abstract{
We perform lattice simulations of two flavors QCD using the optimal domain-wall fermion,  
in which the chiral symmetry is preserved to a good precision ($ m_{res} \sim 0.3 $ MeV) 
on the $ 16^3 \times 32 $ lattice ($ L \sim 2 $~fm) with 
inverse lattice spacing $ a^{-1} \sim 1.8 $~GeV, 
and $ N_s = 16 $ in the fifth dimension,    
for eight sea quark masses corresponding to the pion masses in the range 210-500~MeV.  
We present our first results of the mass and the decay constant of the pseudoscalar meson,  
which are in good agreement with the next-to-leading order chiral perturbation theory   
for $ M_\pi < 450 $~MeV, and from which we determine the low-energy constants   
$ f $, $ \Sigma $, $ \bar{l}_3 $ and $ \bar{l}_4 $. 
At the physical pion mass $ M_{\pi} = 135$~MeV, we
obtain the pion decay constant $f_\pi =133(1)(2)$~MeV, 
and the average up and down quark mass 
$m_{ud}^{\overline{\rm MS}}(\mathrm{2~GeV})=4.09(7)(11)$~MeV, 
where the first error is statistical, 
and the second error is systematic due to the truncation of the higher
order corrections and the uncertainty in the determination of the lattice spacing.
Furthermore, we also obtain the chiral condensate   
$ \Sigma^{\overline{{\mathrm{MS}}}}(\mbox{2~GeV}) = [\mbox{250(4)(7)~MeV}]^3 $.
}

\FullConference{The XXVIII International Symposium on Lattice Field Theory\\
                June 14-19,2010\\
                Villasimius, Sardinia Italy}

\begin{document}

\section{Introduction}

Lattice QCD with exact chiral symmetry is an ideal theoretical framework to study
the nonperturbative physics from the first principles of QCD.
However, it is rather nontrivial to perform Monte Carlo simulation 
such that the chiral symmetry is perserved to a very high precision 
and all topological sectors are sampled ergodically.

Since 2009, the Taiwan Lattice QCD Collaboration (TWQCD) has been using a
GPU cluster (currently constituting of 250 NVIDIA GPUs)
which attains 40 Teraflops (sustained) to simulate unquenched lattice QCD
with the optimal domain-wall quarks \cite{Chiu:2002ir, Chiu:2009wh}.
We have realized our goal of preserving the chiral symmetry
to a good precision (with $ m_{res} \sim 0.3 $ MeV) and also sampling 
all topological sectors ergodically.

In this paper, we present our first results of the mass and the decay constant 
of the pseudoscalar meson in two flavors QCD, and compare our results with 
the next-to-leading order (NLO) chiral perturbation theory (ChPT). 
We find that our data is in good agreement with NLO ChPT for $ M_\pi $ less than 
450 MeV, and from which we determine 
the low-energy constants $ f $, $ \Sigma $, $ \bar{l}_3 $ and $ \bar{l}_4 $, and
and the average up and down quark mass $m_{ud}^{\overline{\rm MS}}(\mathrm{2~GeV})$.
Our result of the topological susceptibility is presented in Ref. \cite{Hsieh:2010ab}, 
and our strategy of using GPU to speed up our Hybrid Monte Carlo simulations 
is presented in Ref. \cite{Chiu:2010gp}.

\section{Hybrid Monte Carlo Simulation with Optimal Domain-Wall Quarks}

The optimal domain-wall fermion 
is the theoretical framework which preserves the 
(mathematically) maximal chiral symmetry for any finite $N_s$ (the length of the fifth dimension). 
Thus the artifacts due to the chiral symmetry breaking with finite $ N_s $ 
can be reduced to the minimum.

The action of the optimal domain-wall fermion is defined as \cite{Chiu:2002ir} 
\bea
\label{eq:ODWF}
S_\mathrm{odwf}
= \sum_{s,s'=1}^{N_s} \sum_{x,x'} 
  \bar\psi_{xs} \left[ (\omega_s D_w + 1)_{xx'} \delta_{ss'}
                      +(\omega_s D_w - 1)_{xx'} L_{ss'}      \right] \psi_{x's'} 
\equiv \bar\Psi \Dodwf \Psi, 
\eea
where the weights $ \{ \omega_s \} $ along the fifth dimension are 
fixed according to the formula derived in Ref. \cite{Chiu:2002ir} 
such that the maximal chiral symmetry is attained. 
Here $D_w$ denotes the standard Wilson-Dirac operator plus a negative parameter $-m_0\; (0 < m_0 < 2)$,
\begin{equation}
(D_w)_{xx'} = -\frac{1}{2} \sum_{\mu} \left[
  (1-\gamma_\mu)U_\mu(x)\delta_{x+\hat{\mu},x'}
 +(1+\gamma_\mu)U^\dagger_\mu(x')\delta_{x-\hat{\mu},x'} \right]
 + (4 - m_0),
\end{equation}
and 
\begin{equation}
L = P_+ L_+ + P_- L_-, \quad P_\pm = (1\pm \gamma_5)/2,
\end{equation}
\begin{equation}
(L_+)_{ss'} = \left\{ 
    \begin{array}{ll} \delta_{s-1,s'}, & 1 < s \leq N_s \\ 
        -(m_q/2m_0) \delta_{N_s,s'}, & s = 1 \end{array}\right.;
\quad\quad L_-=(L_+)^T, 
\end{equation}
where $ m_q $ denotes the bare quark mass.  
Separating the even and the odd sites on the 4D space-time lattice,
$ \Dodwf $ can be written as
\begin{equation}
\Dodwf(m_q)=
S_1^{-1}
\begin{pmatrix}
1 & M_5 D_w^{\text{EO}}  \\
M_5 D_w^{\text{OE}} & 1 
\end{pmatrix}
S_2^{-1}
=S_1^{-1}
\begin{pmatrix}
1 & 0 \\
M_5 D_w^{\text{OE}} & 1
\end{pmatrix}
\begin{pmatrix}
1 & 0 \\
0 & C
\end{pmatrix}
\begin{pmatrix}
1 & M_5 D_w^{\text{EO}} \\
0 & 1
\end{pmatrix}
S_2^{-1},
\label{eq:D_odwf_decomp}
\end{equation}
where 
\begin{equation}
\label{eq:m5}
M_5\equiv \left[(4-m_0) + \sqrt{\omega}^{-1}(1-L)(1+L)^{-1}\sqrt{\omega}^{-1}\right]^{-1},
\quad (\omega)_{ss'} = \omega_s \delta_{ss'}, 
\end{equation}
\begin{equation}
S_1\equiv M_5 \sqrt{\omega}^{-1}, \quad S_2\equiv (1 +  L)^{-1} \sqrt{\omega}^{-1}, 
\end{equation}
and the Schur decomposition has been used in the last equality of (\ref{eq:D_odwf_decomp}), 
with the Schur compliment 
\begin{equation}
\label{eq:c_def}
C = 1 - M_5 D_w^{\text{OE}} M_5 D_w^{\text{EO}}. 
\end{equation}
Since $ \det\Dodwf = \det S_1^{-1} \cdot \det C \cdot \det S_2^{-1} $, and
$ S_1 $ and $ S_2 $ do not depend on the gauge field, we can just use $ C $
for the HMC simulation. After including the Pauli-Villars fields (with $ m_q = 2 m_0 $), the pseudo-fermion
action for 2-flavor QCD ($ m_u = m_d $) can be written as
\bea
\label{eq:Spf}
S_{pf} = \phi^\dagger C_{PV}^\dagger ( C C^\dagger)^{-1} C_{PV} \phi, \quad C_{PV} \equiv C(2m_0).
\eea

In the HMC simulation, we first generate random noise vector $ \xi $ with Gaussian distribution,
then we obtain $ \phi = C_{PV}^{-1} C \xi $ using the conjugate gradient (CG).
With fixed $ \phi $, the system is evolved with a fictituous Hamiltonian dynamics,
the so-called molecular dynamics (MD). In the MD, we use the Omelyan integrator \cite{Takaishi:2005tz},
and the Sexton-Weingarten multiple-time scale method \cite{Sexton:1992nu}.
The most time-consuming part in the MD is to compute the vector $ \eta = (C C^\dagger)^{-1} C_{PV} \phi $
with CG, which is required for the evaluation of the fermion force in the equation
of motion of the conjugate momentum of the gauge field. Thus we program the GPU 
to compute $ \eta $, using CG with mixed precision \cite{Chiu:2010gp}.
Also, we have ported the computation of the gauge force and the update of the gauge field to the GPU.

Furthermore, we introduce an extra heavy fermion field with mass $ m_H $ ($ m_q \ll m_H < 2 m_0 $), 
similar to the case of the Wilson fermion \cite{Hasenbusch:2001ne}.
For two flavors QCD, the pseudofermion action becomes
\bea
S_{pf}^H = \phi^{\dagger} C_H^{\dagger} \frac{1}{CC^{\dagger}} C_H \phi + 
           \phi_H^{\dagger} C_{PV}^{\dagger} \frac{1}{C_H C_H^{\dagger}} C_{PV} \phi_H, \quad C_H \equiv C(m_H),  
\eea
which gives exactly the same fermion determinant of (\ref{eq:Spf}).   
However, the presence of the heavy fermion field plays the crucial role in reducing
the light fermion force and its fluctuation, thus diminishes the change of the Hamiltonian 
in the MD trajactory, and enhances the acceptance rate.

For a system with CPU and GPU, we can have both of them compute concurrently, e.g.,
while the GPU is working on the CG of the light quark field, the CPU can compute the
fermion force of the heavy fermion field. This asynchronous concurrent excecution mode
enhances the overall performance by $\sim 5\% $.

A detailed description of our HMC simulations will be 
presented in a forthcoming paper \cite{Chiu:HMC}.

\begin{figure}[htb]
\centerline{\includegraphics[width=90mm,clip=true]{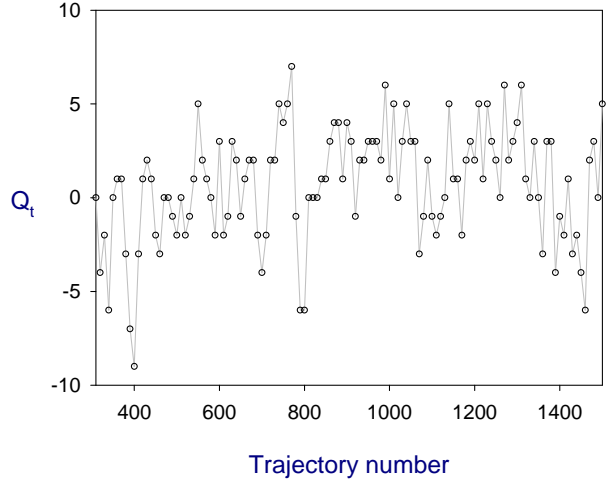}}
\caption{
\label{fig:Qt_history}
The topological charge versus the trajectory in the HMC simulation of two flavors QCD with ODWF.
The lattice is $ 16^3 \times 32 $ with the spatial box $ \sim (\mbox{2 fm})^3 $, and the quark mass 
corresponding to $ M_\pi \sim 300 $ MeV. The topological charge is obtained by projecting the zero modes
of the overlap Dirac operator.}
\end{figure}

\section{Lattice setup}

We simulate two flavors ($N_f=2$) QCD on the $16^3 \times 32$
lattice at the lattice spacing $a \sim $ 0.11~fm, 
for eight sea quark masses in the range $ m_q a =0.01, 0.02, \cdots, 0.08 $.
For the gluon part, we use the plaquette action at $\beta$ = 5.90. 
For the quark part, we use the optimal domain-wall fermion with $ N_s = 16 $.
After discarding 300 trajectories for thermalization, we accumulated about
$ 3000-3200 $ trajectories in total for each sea quark mass.
From the saturation of the error (by binning) of the plaquette, as well as
the evolution of the topological charge (see Fig.~\ref{fig:Qt_history}), 
we estimate the autocorrelation time to be $\sim 10 $ trajectories. 
Thus we sample one configuration every 10 trajectories. 
Then we have $ 270-290 $ configurations for each sea quark mass.

\begin{figure}[htb]
\centerline{\includegraphics[width=90mm,clip=true]{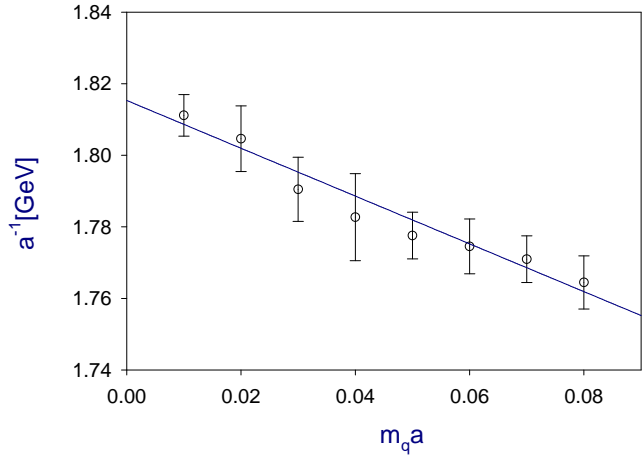}}
\caption{
\label{fig:aimq}
The inverse lattice spacing $ a^{-1} $~[GeV] versus $ m_q a $ for two flavors QCD with ODWF.}
\end{figure}

We determine the lattice spacing by heavy quark potential with Sommer parameter $ r_0 = 0.49 $ fm.
The inverse lattice spacing versus the quark mass is plotted in Fig.~\ref{fig:aimq}. 
Using the linear fit, we obtain the inverse lattice spacing in the chiral limit, 
$ a^{-1} = 1.8153(28)$ GeV. 

For each configuration, we calculate the exact zero modes plus
80 conjugate pairs of the lowest-lying eignmodes of the overlap Dirac operator.
We outline our procedures as follows.
First, we project 240 low-lying eigenmodes of $ H_w^2 $ using $\nu$-TRLan
alogorithm \cite{nu-TRLan}, where each eigenmode has a residual less than $ 10^{-12} $.
Then we approximate the sign function of the overlap operator
by the Zolotarev optimal rational approximation with 64 poles,
where the coefficents are fixed with $ \lambda_{max}^2 = (6.4)^2 $,
and $ \lambda_{min}^2 $ equal to the maximum of
the 240 projected eigenvalues of $ H_w^2 $.
Then the sign function error is less than $ 10^{-14} $.
Using the 240 low-modes of $ H_w^2 $ and the Zolotarev approximation
with 64 poles, we project the zero modes plus 80 conjugate pairs of
the lowest-lying eignmodes of the overlap operator
with the $\nu$-TRLan algorithm,
where each eigenmode has a residual less than $ 10^{-12} $.

We measure the chiral symmetry breaking (due to finite $N_s$) by computing the residual mass
\bea
\label{eq:mres}
m_{res} \equiv \left< \frac{\sum_x \left< J_5(x,N_s) \bar q(0) \gamma_5 q(0) \right>}
                      {\sum_x \left< \bar q(x) \gamma_5 q(x) \bar q(0) \gamma_5 q(0) \right>} \right>_{\{U\}}
= \left< \frac{ \tr(D_c + m_q)^{-1}_{0,0} }{ \tr[(D_c^\dagger + m_q)(D_c+m_q)]^{-1}_{0,0} } \right>_{\{U\}} - m_q,   
\eea
where $ (D_c + m_q)^{-1} $ is the valence quark propagator with $ m_q $ equal to the mass of the sea quark, 
tr denotes the trace running over the color and Dirac indices, and the subscript $ \{U\} $ denotes averaging 
over an ensemble of gauge configurations. It turns out that, after averaging over an ensemble of a few 
hundreds of independent gauge configurations, $ m_{res} $ is insensitive to the location of 
the origin $ x^\mu = (0, 0, 0, 0) $. Thus (\ref{eq:mres}) gives a reliable measure of 
chiral symmetry breaking due to finite $ N_s $.
The derivation of (\ref{eq:mres}) will be given in a forthcoming paper \cite{Chen:2011}.

\begin{figure}[htb]
\centerline{\includegraphics[width=100mm,clip=true]{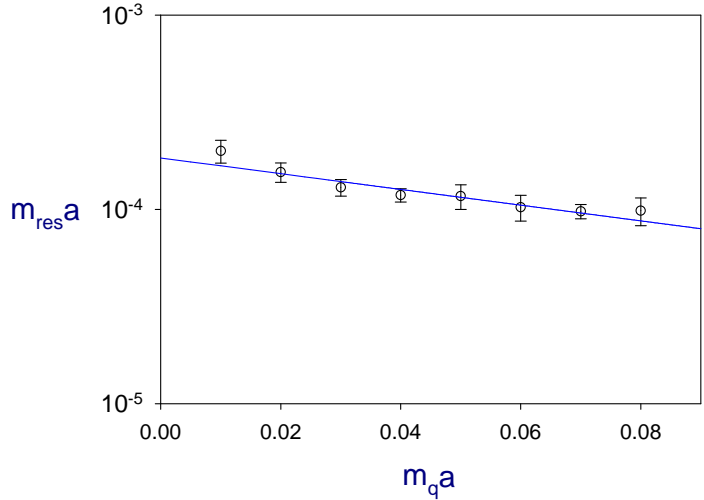}}
\caption{
\label{fig:mres}
The residual mass versus the quark mass for two flavors QCD with ODWF.}
\end{figure}

In Fig. \ref{fig:mres}, we plot the residual mass versus the quark mass.   
Using the power-law fit, we obtain the residual mass in the chiral limit, 
$ m_{res} a = 0.00018(2) $, which amounts to $ m_{res} = 0.32(4) $~MeV.  
Note that the value of $ m_{res} $ is less than 1/10 of the statistical 
and systematic errors of the inverse lattice spacing,   
thus confirming that the chiral symmetry has been preserved 
to a good precision in our simulation.

\section{The Mass and the Decay Constant of the Pseudoscalar Meson}

\begin{figure}[tb]
\begin{center}
\begin{tabular}{@{}c@{}c@{}}
\includegraphics*[width=7.55cm,height=5.8cm]{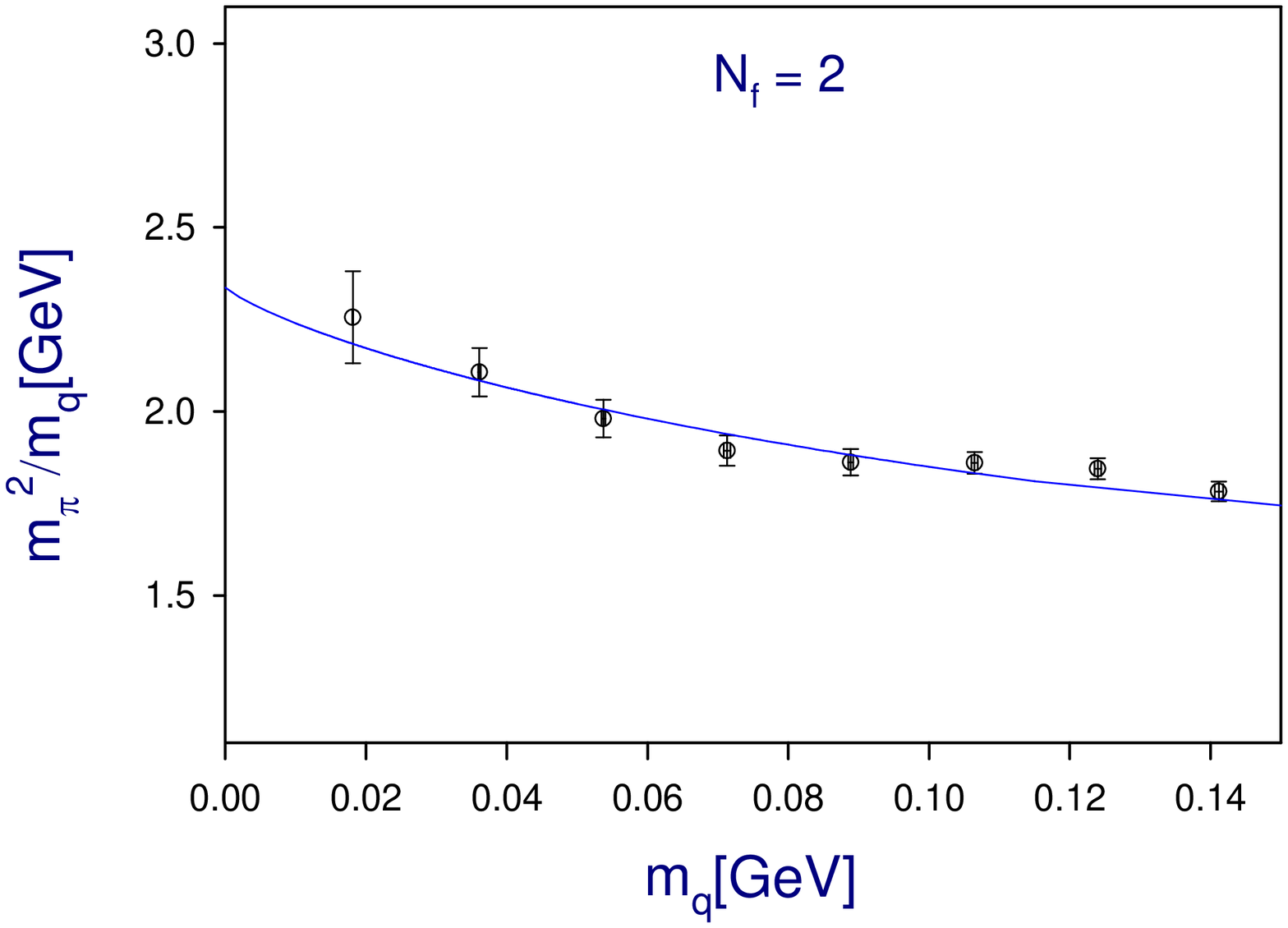}
&
\includegraphics*[width=7.55cm,height=5.8cm]{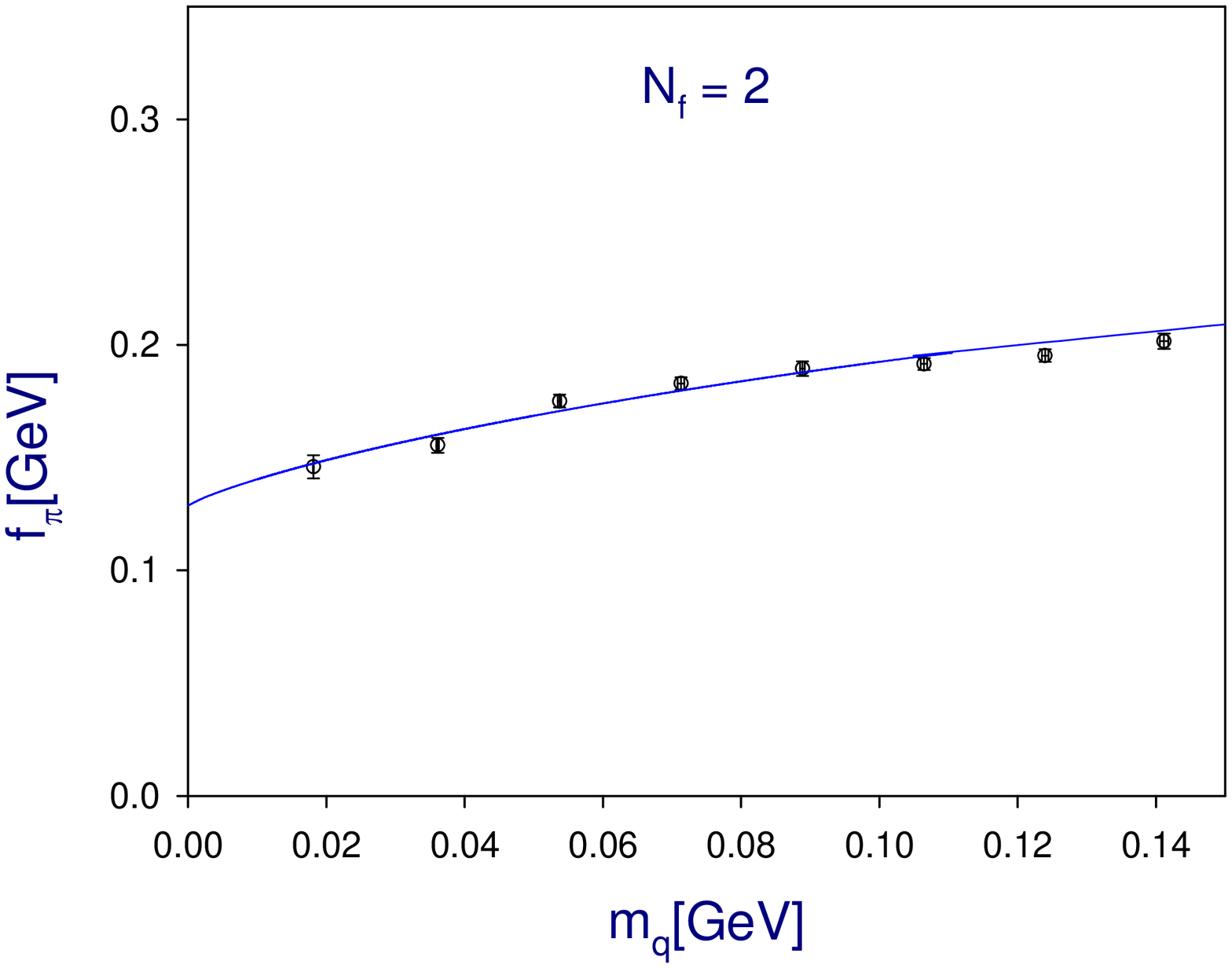}
\\ (a) & (b)
\end{tabular}
\caption{ Physical results of 2-flavor QCD with optimal domain-wall quarks:
          (a) $ m_\pi^2/m_q $, and (b) $ f_\pi $.
          The solid lines are the simultaneous fits to the NLO ChPT, for the six lightest quark masses.}
\label{fig:mpi2omq_fpi_b590_nf2}
\end{center}
\end{figure}

In this section, we present our first results of the pseudoscalar mass and decay constant,
for 2 flavors QCD with optimal domain-wall quarks and the plaquette gluon action at 
$ \beta = 5.90 $, on the $ 16^3 \times 32 \times 16 $ lattice.
In Fig. \ref{fig:mpi2omq_fpi_b590_nf2}, we plot $ M_\pi^2 /m_q $ and $ f_\pi $ versus $ m_q $
respectively. Here we have made the correction for the finite volume effect 
using the estimate within ChPT calculated up 
to $ {\cal O}(M_\pi^4/(4\pi f_\pi)^2 ) $ \cite{Colangelo:2005gd}, 
since our simulation is done on a finite volume
lattice with $ M_\pi L \sim 2.0 $ for the lightest sea quark, and 
its finite volume effect cannot be neglected. 
 
Taking into account of the correlation between $ M_\pi^2/m_q $ and $ f_\pi $ for the same sea quark mass,  
we fit our data to the formulas of the next-to-leading order (NLO) chiral perturbation theory (ChPT) \cite{Gasser:1984gg}
\bea
\label{eq:mpi2omq_NLO_Nf2}
\frac{M_\pi^2}{m_q} &=& 2 B \left[ 1 
+ \left(\frac{2 B m_q }{16 \pi^2 f^2}\right) \ln\left(\frac{2 B m_q}{\Lambda_3^2} \right) \right], \quad B \equiv \frac{2 \Sigma}{f^2} \\
\label{eq:fpi_NLO_Nf2}
f_\pi &=& f \left[ 1 -  \left(\frac{4 B m_q}{16 \pi^2 f^2 } \right) \ln \left( \frac{2 B m_q}{\Lambda_4^2} \right) \right], 
\eea
where $ \Lambda_i $ are related to the low energy constants $ \bar l_i $  
\bea
\bar l_3 = \ln \left( \frac{\Lambda_3^2}{m_{\pi^{\pm}}^2} \right), \quad 
\bar l_4 = \ln \left( \frac{\Lambda_4^2}{m_{\pi^{\pm}}^2} \right), \quad m_\pi^{\pm} = 0.140 \mbox{ GeV}.  
\eea
 
For the six lightest quark masses (corresponding to pion masses in the range $210-445$ MeV), 
our fit gives 
\bea
\label{eq:Sfl3l4}
\Sigma = 0.2105(30) \mbox{ GeV}, \quad  
f = 0.127(2) \mbox{ GeV}, \quad 
\bar l_3 = 4.37(18), \quad 
\bar l_4 = 5.31(11),  
\eea   
with $ \chi^2$/dof = 0.4. 
At the physical pion mass $ M_\pi \simeq 0.135 $ GeV, 
the value of pion decay constant is $ f_\pi = 0.133(1) $ GeV,  
and the bare quark mass is $ 0.0069(2) $ GeV. 
In order to convert the bare quark mass to that in the
$\overline{\mathrm{MS}}$ scheme, we calculate the
renormalization factor $Z_m^{\overline{\mathrm{MS}}}(\mathrm{2~GeV})$
using the non-perturbative renormalization technique
through the RI/MOM scheme \cite{Martinelli:1994ty}, and 
obtain $Z_m^{\overline{{\mathrm{MS}}}}(\mbox{2 GeV}) = 0.5934(10)$ \cite{Chiu:NPR}.
Then the value of the average up and down quark mass is transcribed to
\bea
\label{eq:mudMS}
m_{ud}^{\overline{{\mathrm{MS}}}}(\mbox{2 GeV}) = 4.09(7)(11) \mbox{ MeV}, 
\eea
Similarly, the value of $ \Sigma $ in (\ref{eq:Sfl3l4}) is transcribed to
\bea
\label{eq:sigmaMS}
\Sigma^{\overline{{\mathrm{MS}}}}(\mbox{2 GeV}) = [250(4)(7) \mbox{ MeV}]^3 
\eea
The systematic error is estimated from the turncation of higher order 
effects and the uncertainty in the determination of lattice spacing with 
$ r_0 = 0.49 $ fm.
Since our calculation is done at a single lattice spacing,
the discretization error cannot be quantified reliably, but
we do not expect much larger error because our lattice
action is free from $O(a)$ discretization effects.

\section{Concluding remarks}

Using a GPU cluster (currently attaining 40 sustained Teraflops with
250 NVIDIA GPUs), we have succeeded to simulate unquenched lattice QCD
with optimal domain-wall quarks, which preserves the chiral symmetry
to a good precision 
and samples all topological sectors ergodically. 
Our results of the mass and the decay constant
of the pseudoscalar meson (in this paper) 
and the topological susceptibility (in Ref. \cite{Hsieh:2010ab})
suggest that the nonperturbative chiral dynamics of the sea quarks are well under 
control in our simulations. This provides a new strategy to tackle QCD nonperturbatively 
from the first principles.

\begin{acknowledgments}
  This work is supported in part by the National Science Council
  (Nos.~NSC96-2112-M-002-020-MY3, NSC99-2112-M-002-012-MY3, NSC96-2112-M-001-017-MY3, 
  NSC99-2112-M-001-014-MY3, NSC99-2119-M-002-001) and NTU-CQSE~(Nos.~99R80869, 99R80873).
  We are grateful to NCHC and NTU-CC for providing facilities to perform some of the computations.  
  We also thank Kenji Ogawa for his contribution in the development of simulation code.

\end{acknowledgments}

\end{document}